# Material Approximation of Data Smoothing and Spline Curves Inspired by Slime Mould

Jeff Jones and Andrew Adamatzky

Centre for Unconventional Computing
University of the West of England
Coldharbour Lane
Bristol, BS16 1QY, UK.
`jeff.jones@uwe.ac.uk, andrew.adamatzky@uwe.ac.uk`

**Abstract.** The giant single-celled slime mould *Physarum polycephalum* is known to approximate a number of network problems via growth and adaptation of its protoplasmic transport network and can serve as an inspiration towards unconventional, material-based computation. In *Physarum* predictable morphological adaptation is prevented by its adhesion to the underlying substrate. We investigate what possible computations could be achieved if these limitations were removed and the organism was free to completely adapt its morphology in response to changing stimuli. Using a particle model of *Physarum* displaying emergent morphological adaptation behaviour we demonstrate how a minimal approach to collective material computation may be used to transform and summarise properties of spatially represented datasets. We find that the virtual material relaxes more strongly to high-frequency changes in data which can be used for the smoothing (or filtering) of data by approximating moving average and low-pass filters in 1D datasets. The relaxation and minimisation properties of the model enable the spatial computation of B-spline curves (approximating splines) in 2D datasets. Both clamped and unclamped spline curves, of open and closed shapes, can be represented and the degree of spline curvature corresponds to the relaxation time of the material. The material computation of spline curves also includes novel quasi-mechanical properties including unwinding of the shape between control points and a preferential adhesion to longer, straighter paths. Interpolating splines could not directly be approximated due to the formation and evolution of Steiner points at narrow vertices, but were approximated after rectilinear pre-processing of the source data. This pre-processing was further simplified by transforming the original data to contain the material inside the polyline. These exemplar results expand the repertoire of spatially represented unconventional computing devices by demonstrating a simple, collective and distributed approach to data and curve smoothing.

# 1 Introduction - Morphological Computation

Morphological computation seeks to directly utilise the embodied properties of synthetic systems and their interactions with the environment in which they exist [33]. The motivations for this approach are twofold. The first is based on necessity caused by the increasing complexity of computing systems. Traditional approaches in integrating separate complex systems (for instance, sensors, actuators and control systems in robotics) result in complex components with little redundancy or fault tolerance. It is typically difficult to control such systems as the size of the system increases. The second motivation has more positive origins. Can we utilise the natural, material properties of synthetic systems to improve the performance (or minimise the difficulties faced) when constructing synthetic systems? Furthermore, can we extend the properties of embodiment to include the environment and take advantages of the processes which occur naturally within it?

We can gain some help in investigating approaches to morphological computation by examining the natural world. Nature inspired computing and robotics takes inspiration from systems which are composed of a great many relatively simple parts and which are embedded within complex environments. For example, swarm computation approaches seek to elucidate the sensory mechanisms and individual interactions which generate the complexity patterning and movement seen at very different scales in natural systems including car traffic dynamics [14], human walking patterns [15], flocking and schooling [36], collective insect movement [5], and bacterial patterning [28]. In all these examples there is a population of entities in space, coupled by sensory information about their environment. The collective morphology of the group is generated from the local interactions and movement of individual members of the population. These interactions generate complex, self-organised and emergent behaviour. The resulting morphological patterns of the population show adaptation to the environment, for example avoidance of obstacles, detection of prey, and avoidance of predators.

The tenets of emergent behaviour (simple, local interactions with self-organisation) may also be exhibited in simpler systems which straddle the boundary of non-living physical materials and living organisms including those acting as biological fibres and membranes [44], lipid self assembly in terms of networks [26], pseudopodium-like membrane extension [27] and even those exhibiting simple chemotaxis responses [25]. Some engineering and biological insights have already been gained by studying the structure and function of what might be termed 'semi-biological' materials and the complex behaviour seen in such minimal examples raises questions about the lower bounds necessary for the emergence of apparently intelligent behaviour. It has also been suggested that simple material behaviour, as opposed to complex living systems, may provide a rich vein of potential computational resources to be explored [40].

## 1.1 Morphological Computation as Performed by Slime Mould

The above examples of collective morphological adaptation in biological and semi-biological systems suggest a literal form of morphological computation, driven by dynamic patterning and responding to a complex environment. An ideal hypothetical candidate for a morphological computation medium would be a material which is capable of the complex sensory integration, movement and adaptation of a living organism, yet which is also composed of relatively simple components that are amenable to understanding and control. The Myxomycete organism, the true slime mould *Physarum polycephalum*, may be a suitable candidate medium which meets both criteria; i.e. it exhibits complex behaviour, but which is composed of relatively simple materials.

A giant single-celled organism, *Physarum* is an attractive candidate medium for morphological computation because it is a literal example of morphological adaptation. During the plasmodium stage of its life cycle it adapts its body plan in response to a range of environmental stimuli (nutrient attractants, repellents, hazards). Fig. 1,a shows a plasmodium which was inoculated on an oat flake at the left of the figure. The plasmodium is attracted by the chemoattractant gradient from the flake on the right side and streams towards the stimulus. Behind the active growth zone (Fig. 1,a right side), the plasmodium forms a dense network of protoplasmic tubes which coarsens over time (the single tube in Fig. 1,a which connects the two flakes). This tube network is used to distribute nutrients within the plasmodium. The organism is remarkable in that its complex behaviour is achieved without any specialised nervous tissue. Control of its shape and behaviour is distributed throughout the simple material comprising the cell and the cell can survive damage, excision or even fusion with another cell.

The plasmodium of slime mould is amorphous in shape and ranges from the microscopic scale to over a square metre in size. It is a giant single-celled syncytium formed by repeated nuclear division, comprised of a sponge-like actomyosin complex co-occurring in two physical phases. The gel phase is a dense matrix subject to spontaneous contraction and relaxation, under the influence of changing concentrations of intracellular chemicals. The protoplasmic sol phase is transported through the plasmodium by the force generated by the oscillatory contractions within the gel matrix. An example of plasmodium growth and Protoplasmic flux, and thus the behaviour of the organism, is affected by changes in pressure, temperature, space availability, chemoattractant stimuli and illumination [6], [9], [24], [29], [31], [41], [43]. The *Physarum* plasmodium can thus be regarded as a complex functional material capable of both sensory and motor behaviour. Indeed *Physarum* has been described as a membrane bound reaction-diffusion system in reference to both the complex interactions within the plasmodium and the rich computational potential afforded by its material properties [3]. The study of the computational potential of the *Physarum* plasmodium was initiated by Nakagaki et al. [30] who found that the plasmodium could solve simple maze puzzles. This research has been extended and the plasmodium has demonstrated its performance in, for example, path planning and plane division problems [39],[38], spanning trees and proximity graphs [2], [1], simple mem-ory effects [37], the implementation of individual logic gates [42] and *Physarum* inspired models of simple adding circuits [20].

### 1.2 Computational Limitations of Slime Mould

Although *Physarum* slime mould has desirable computational properties, it also has some practical limitations. Although relatively simple and inexpensive to culture, its computation is slow, taking many hours — or even days — during which time it must be maintained within strict environmental parameters of temperature, light exposure and humidity. *Physarum* may also be relatively unpredictable in its behaviour which, although useful in wild conditions, is a hindrance when repeatability is concerned.

The protoplasmic transport networks formed by the plasmodium are not as precise as those constructed using classical approaches. The network paths meander and branch between the nutrient nodes (Fig. 1, a) and do not necessarily pass directly through the nodes (Fig. 1, b), giving an imprecise result. A growing plasmodium also produces a slime capsule surrounding the organism which anchors the plasmodium to its substrate, limiting the scope for further adaptation and minimisation of the tube network (greater network adaptation can be seen when growing the organism on a water substrate [4] but this is impractical for most purposes). *Physarum* responds to the adhesion by either spawning new tube growth from existing tubes, or abandoning unused tubes. However, the remnants of the old tubes still remain (Fig. 1, c), and influence future movement choices of the plasmodium [34]. The presence of imprecise network paths, old tube remnants, extraneous tubes, or tortuous tube paths, can prevent a clear result from being displayed. Finally, in classical computation, problems and their solutions may be represented symbolically. However, in *Physarum* computing the problem, and solution, must be given a direct spatial representation. Although this is not a limitation *Per se*, it does require (perhaps unfamiliar) techniques of representing problems in space to exploit the distributed and parallel computational properties of the organism.

### 1.3 Towards Modelling Morphological Adaptation in Slime Mould

Due to the practical and computational limitation of slime mould we require a synthetic analogue of the organism to explore possibilities of spatially represented computation. One technique available is computer modelling, where we attempt to reproduce the complex patterning of slime mould along with the complex interactions it has with its environment. However, we are not simply trying to extract the features of slime mould for classical algorithms. Such an approach may indeed prove useful for certain tasks, but would not inform us in any way about the distributed emergent behaviour and control of the organism. Instead we wish to construct a virtual material using the same principles of slime mould. Namely, simple component parts and local interactions. The aim is to generate collective emergent behaviour utilising self-organisation to yield an embodied form of material computation which can reproduce the wide range of complex patterning and environmental responses seen in slime mould, without some of the drawbacks caused by the particular living and physical properties of the organism itself.

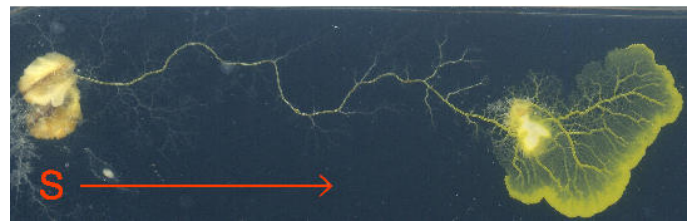

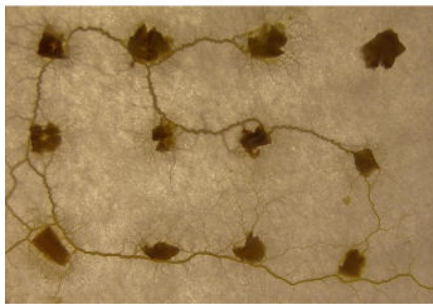

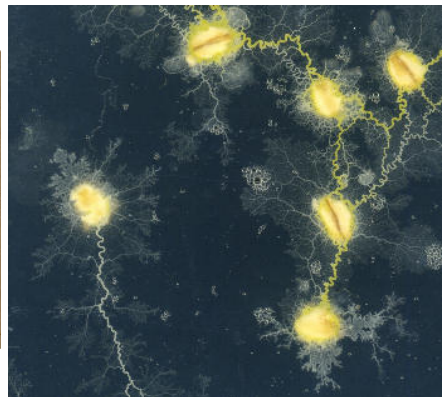

**Fig. 1.** *Physarum* tube networks networks and features which restrict computational functionality. a) *Physarum* plasmodium on plain agar inoculated on the left oat flake (S) grows towards the flake on the right, forming a protoplasmic tube connecting the nutrients, b) protoplasmic tube network connecting multiple nutrient oat flakes (damp filter paper substrate, backlit illumination), c) slime capsule and remnants of previous vacant tubes (paler paths) act to constrain network adaptation and path choice.

## 2 Basic Method and Overview of Paper

We used the multi-agent approach introduced in [18], in which we introduced a large population of simple mobile agents whose behaviour was coupled via a diffusive chemoattractant lattice. Agents sense the concentration of a hypothetical 'chemical' in the lattice, orient themselves towards the locally strongest source and deposit the same chemical during forward movement. The collective movement trails spontaneously form emergent transport networks which undergo complex evolution, exhibiting minimisation and cohesion effects under a range of sensory parameter settings. The collective behaves as a virtual material demonstrating characteristic network evolution motifs and minimisation phenomena seen in soap film evolution (for example the formation of Plateau angles, T1 and T2 relaxation processes and adherence to von Neumann's law). A full exploration of the dynamical patterns were explored in [17] which found that the population could reproduce a wide range of Turing-type reaction-diffusion patterning. Nutrients (representing data points) are represented by projecting chemoattractants into the lattice at fixed sites and the network evolution is constrained by the distribution of nutrients. Network evolution is affected by nutrient distribution and nutrient concentration. It was found that the emergent transport networks reproduced the connectivity of slime mould by approximating networks in the Toussaint hierarchy of proximity graphs [19], as originally demonstrated in [1]. Using a combination of attractant and repellent stimuli the method is also capable of approximating the Convex Hull, Concave Hull and Voronoi diagram [22].

In this paper we utilise the morphological adaptation inherent in the model to explore the effect of complete adaptation of a virtual material to spatial datasets. Specifically we explore the approximation of data smoothing functions and spline curves. We use model parameter settings which ensure strong adaptation of a pre-existing network and a full description of the model is given in the appendix. In Section 3 we examine the properties of the material adaptation in response to high and low frequency data variations and give examples showing the potential to summarise simple statistical properties of data in 1D space for smoothing functions, including moving average and low pass filtering. In Section 4 we explore 2D datasets and approximate the formation and evolution of spline curves, including B-splines (approximating splines), both unclamped and clamped, in open and closed curves of differing curve degree. Some quasi-mechanical properties of material-based spline curves are noted in Section 5, along with some practical limitations of the approach. In Section 6 we examine the problem of approximating interpolating spline curves using the model and find that interpolating splines are not directly possible due to the presence of narrow vertex angles which cause additional nodes to form in the network. Pre-processing of the original dataset polyline was attempted to resolve this limitation. Pre-processing the original polyline into a rectilinear configuration is explored in Section 6.1 and a simpler method of pre-processing is described in 6.2 in which the polyline is transformed into a wider tube in which the virtual material is then initialised and contained. In Section 7 we summarise the material approach noting some of its useful properties, some limitations and suggest scope for further research and potential applications. Due to the dynamical behaviour of the material adaptation the reader is encouraged to refer to the supplementary video recordings at http://uncomp.uwe.ac.uk/jeff/splines_approximation.htm.

## 3 Material Approximation of Data Smoothing and Filtering

Smoothing and filtering of datasets plays an important factor in all aspects of information technology and signal processing in general. The methods can be as simple as those achieved with a small number of simple analogue electrical components to remove certain aspects of a signal. At the other extreme are complex digital algorithms to enhance certain trends within complex datasets. Can the innate morphological adaptation behaviour of the *Physarum* model be used for data smoothing applications? To test this idea we represented a 1D signal as a sequence of Y-axis data values along a time-line represented by X-axis values (Fig. 2a). The virtual material was patterned in this initial shape and was initially held in place by projecting attractants in the shape of the original pattern (Fig. 2b). We examined the effect of completely removing the attractant stimulus, and the effect of reducing the stimulus strength. When the attractant stimulus was removed the material relaxes and adapts into a profile which smooths the data, matching the moving average of the original data. The moving average is computed conventionally by a kernel computing the mean of the current data point and its left and right data points of window size $w/2$. The moving average filter has non-periodic boundary conditions and subsequently the moving average line narrows as the kernel window size increases (to prevent the window exceeding the bounds of the data). The amount of smoothing by the material deformation (i.e. corresponding to the kernel window size of the moving average) is dependent on the length of time that the material adapts for (Fig. 2c-h). Note that the width of the material also shrinks over time, mirroring the narrowing of the moving average line as the kernel window increases.

Instead of completely removing the original data stimuli it is possible to try and use the original stimuli to constrain the adaptation of the model. This is achieved by projecting a weaker representation of the original data stimuli. The effect of maintaining a weakened stimulus strength was subtly different to the moving average and appeared to partially filter the data. Specifically, the material tended to adhere in areas of the data that were relatively unchanging, whilst detaching from regions that underwent large changes in direction (Fig. 3). This behaviour corresponds to that of an iterative low-pass filtering process where high frequency signal components are removed whilst low frequency components remain. Note that by maintaining a weak stimulus, the width of the material spanning the dataset is not significantly reduced, in contrast to the material approximation of the moving average.

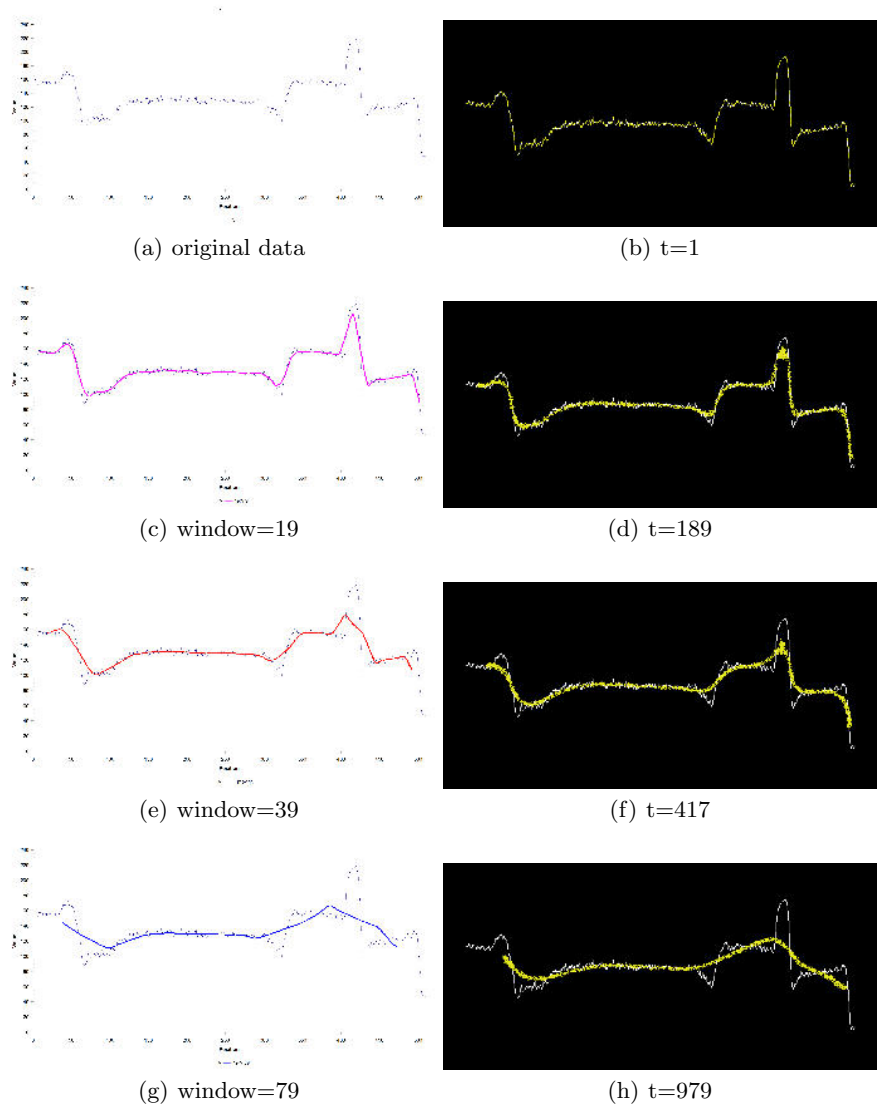

**Fig. 2.** Relaxation of virtual material approximates the moving average. (a,c,e,g) original data (thin line) and overlaid moving average filtered data (thick line) with 1D kernel of size 19, 39 and 79 respectively, (b,d,f,h) initialisation of virtual material on original data followed by snapshots at increasing time intervals.

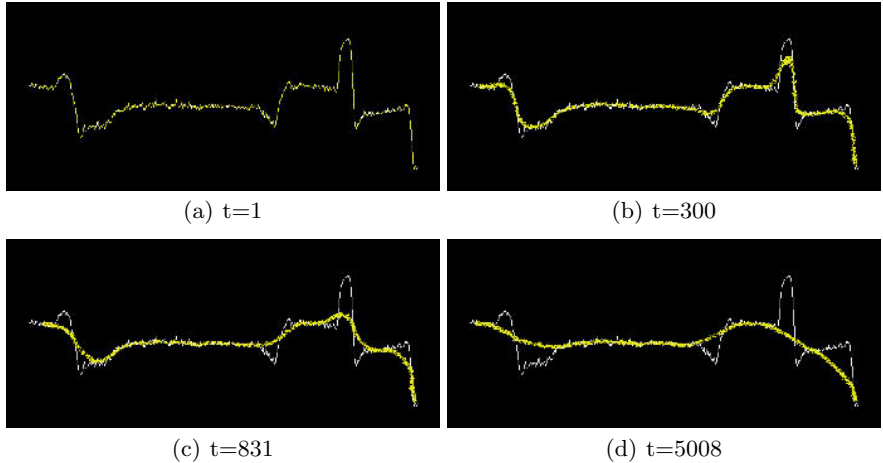

**Fig. 3.** Weak background stimulus constrains adaptation of the virtual material approximating a low pass filter. a) Initialisation of virtual material on original data weakly projected into the lattice at concentration of 0.255 units, b-d) snapshots at increasing time intervals showing removal of sharp peaks and troughs in the data.

In both of these examples the material appears to adapt more strongly to sudden changes in stimuli. To assess this behaviour we initialised the virtual material with the pattern of sine waves and square waves of identical amplitude but different frequencies. Fig. 4 shows snapshots taken which indicate that relaxation occurs more strongly with high frequency changes, for example when comparing Fig. 4 c) and e). The presence of weak stimuli in pattern of the original configuration data appears to act as a brake on the relaxation process. The adhesion of the virtual material to attractant regions results in less curved paths (Fig. 4 d,f) when compared to the smoother paths when no stimulus was present (Fig. 4 c,e).

## 4 Material Approximation of Spline Curves

Splines are mathematical functions constructed piecewise from polynomial functions. Spline functions connect separate data points with a smooth continuous curvature where the individual functions join (at regions called knots) [35],[8]. Spline functions are useful for curve fitting problems (*approximating splines*, where the spline smooths the path between data points) [10]. They may also be used in interpolation problems (*interpolating splines*, where the spline curve passes through all of the data points) [16]. Due to the natural curvature enabled by spline curves, and rapid development in computer aided design systems, they have proven popular in design and architecture [12]. The term spline apparently refers to the use of flexible strips formerly used in the shipbuilding and motor-vehicle industries to allow the shaping of wood and metal shapes into smooth forms by deforming them at selected points using weighted metal objects known as ducks. Thus, there is an inherent mechanical nature to the operation and interpretation of spline curves. The mechanical properties have been used as an inspiration for deformable models and templates, initiated by [23], primarily for image segmentation, and subsequently extended to 3D application [7], [13]. To assess the potential for collective and emergent material shrinkage using the model of *Physarum* we use simpler datasets in these examples.

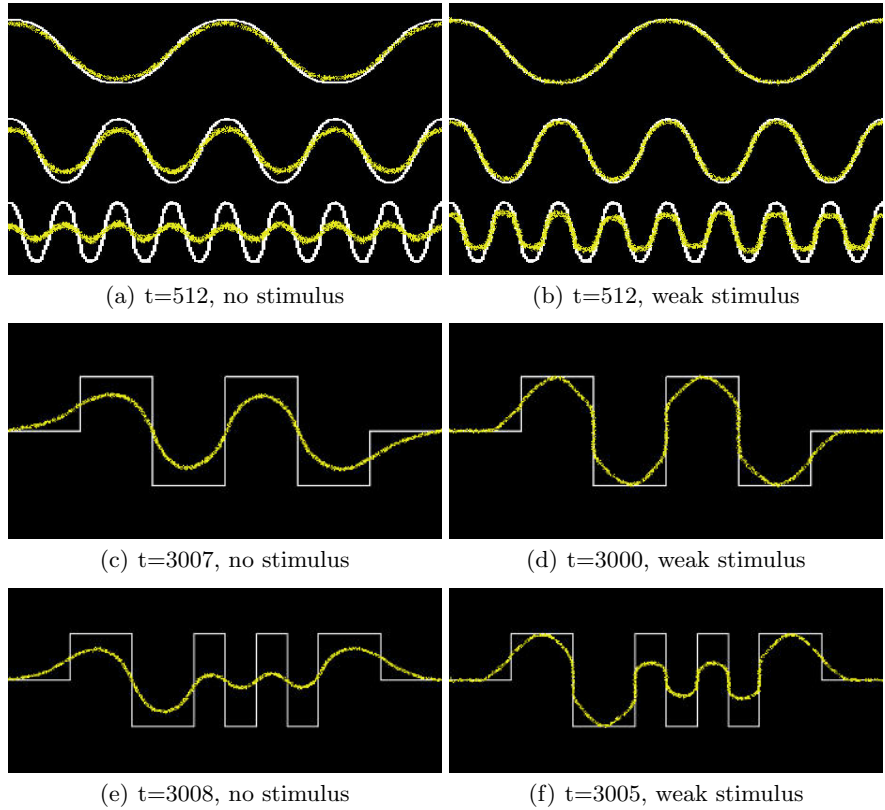

(a) t=512, no stimulus  (b) t=512, weak stimulus

(c) t=3007, no stimulus  (d) t=3000, weak stimulus

(e) t=3008, no stimulus  (f) t=3005, weak stimulus

**Fig. 4.** Differences in relaxation in response to frequency changes and background pattern stimulus. a) original sine wave pattern and relaxation with no stimulus, b) original sine wave pattern and relaxation pattern with weak background stimulus, c) square wave pattern with no stimulus, d) square wave pattern with weak stimulus, e) high frequency square wave with no stimulus, f) high frequency square wave with weak stimulus.

It is known that B-splines (approximating splines) are contained within the Convex-Hull of their original polyline. For this reason it may be a good candidate for approximation by a collective material shrinkage mechanism, since it has previously been shown (for the purposes of combinatorial optimisation) that a large blob of virtual material patterned in the shape of the Convex-Hull adapts and shrinks over time to smaller representations lying within the Convex Hull [21]. We assessed the behaviour of the material under a range of different stimuli conditions to see if it could approximate B-spline curves. In the first case an unclamped quadratic B-spline curve of different degrees is calculated by numerical methods (Fig. 5, a-d). The blob of virtual material was initialised in the pattern of the original polyline connecting the points. During adaptation and shrinkage of the blob without any node stimuli, the pattern approximates the unclamped B-spline. As time progresses the blob adapts the shape of the B-spline of increasing degree (Fig. 5, e-h). Ultimately the blob shrinks to a small point which appears to approximate the centroid position of the original shape. Further investigation of this apparent statistical computation (computation of centroid) is the subject of current research (in preparation).

In the second case a clamped quadratic B-spline curve of different degrees is calculated by numerical methods (Fig. 5, i-l). The virtual material was initialised in the pattern of the original polyline connecting the points and the clamping of the material was enforced by projecting attractant at the location of the two end points. During adaptation and shrinkage of the clamped blob, the pattern approximates the clamped B-spline. As time progresses the blob adapts the shape of the B-spline of increasing degree (Fig. 5, m-p). Adaptation of the unclamped blob occurs more quickly than when the blob is clamped at the control points. The clamped condition shows a non-linear time course and more information on the relationship between curve degree and relaxation time is given in the appendix.

Relaxation of the virtual material may also approximate B-spline curves in more complex configurations, as shown in Fig. 6 in which a set of 20 points is used to generate a B-spline curve of degree 2 and 5. The virtual material, again initialised in the path of the original polyline, relaxes over time when most of the initial attractant stimulus is removed (except for the two end points to clamp the relaxing population). Increasing relaxation time period again corresponds to increasing degree of the original spline curve (Fig. 6 c,d).

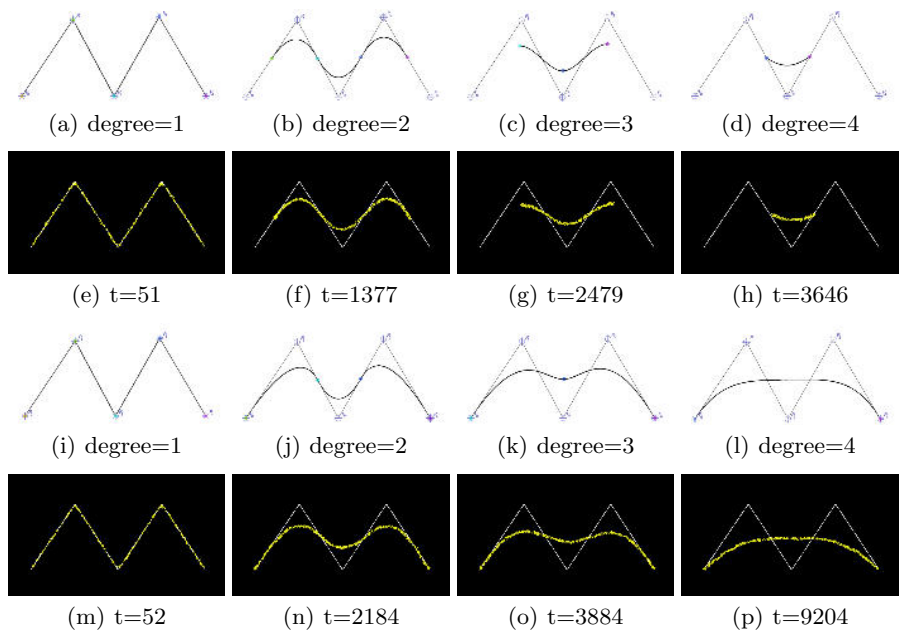

**Fig. 5.** Approximation of B-spline by material adaptation on 'M' shape configuration. Each experiment summarised on a single row. a) unclamped B-spline (5 points connected by polyline) of degree 1, (b-d) same configuration B-splines of degrees 2, 3 and 4 respectively, (e-h) evolution of material adaptation approximates the unclamped B-spline of different degrees, i) clamped B-spline of degree 1 (same configuration as above with start and end points clamped), (j-l) clamped B-spline of degree 2, 3 and 4 respectively, (m-p) evolution of material adaptation approximates the clamped B-spline of different degrees, material is clamped by attractants at end points.

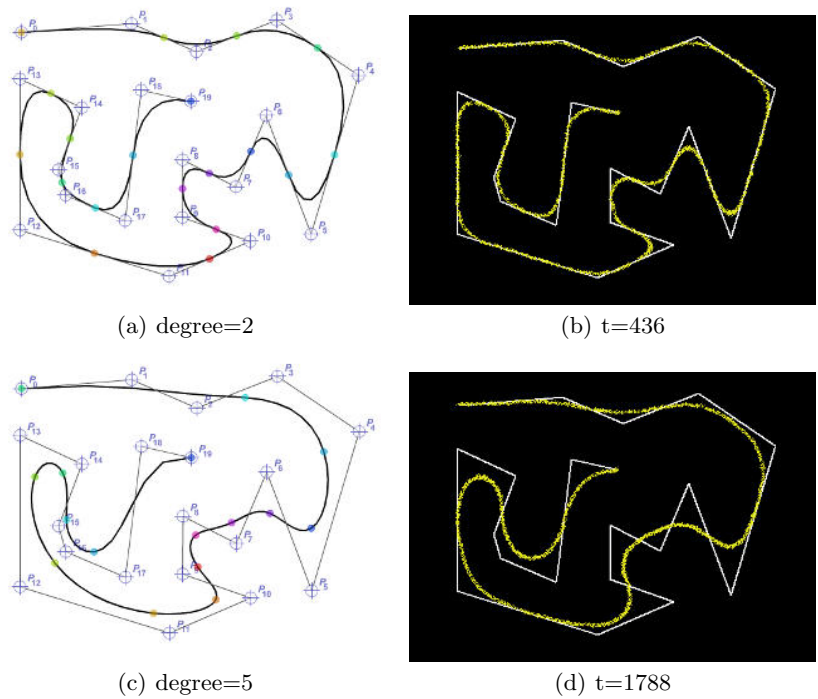

(a) degree=2    (b) t=436

(c) degree=5    (d) t=1788

**Fig. 6.** Clamped B-spline of complex shape and approximation by virtual material. Left: B-spline at differing degrees, composed of 20 points, clamped at start and end points. Points shown as labelled hollow circles connected by faint lines, knots shown as solid circles on thicker spline curve. Right: Approximation of spline curve by relaxation of virtual material, clamped at start and end points.

Closed shapes may be represented by clamped spline curves by repeating the same start data points. For unclamped open shapes, overlapping the first and last three points generates a smooth open curve (Fig. 7,a). For clamped open shapes the first point is overlapped by the end point (Fig. 7,b). For the material approximation of open spline curves the material is simply patterned with the closed polyline (Fig. 7,c) and if a clamping point is required this is represented by attractant projection at the desired clamping site (Fig. 7,d).

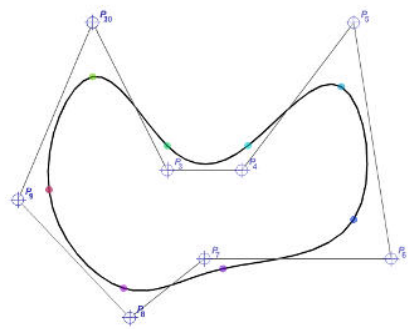
(a) closed shape unclamped B-spline

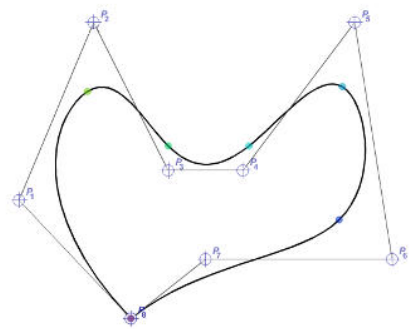
(b) closed shape clamped B-spline

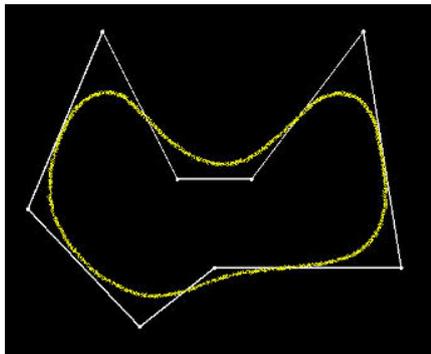
(c) unclamped material t=3880

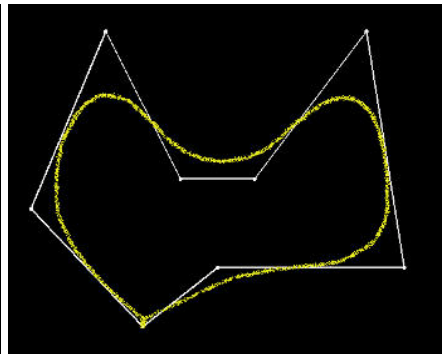
(d) clamped material t=4351

**Fig. 7.** Approximation of B-spline curves in closed cyclic shapes. a) closed shape with no clamping points, b) closed shape clamped at point 8, c) material approximation of unclamped shape, d) material approximation of clamped shape by projecting attractant at data point 8.

# 5 Properties and Limitations of Material-based Spline Curves

## 5.1 Quasi-mechanical Properties

Two unusual properties of the material adaptation approach to spline curves are of note, both relating to the innate mechanical nature of the material adaptation. The first of these effects can be seen by comparing material curves with the classical quadratic B-spline curves of differing degrees. These spline curves retain the proximity of the knots to their respective polylines used in their generation (Fig. 8). The curve formed by the particle population on a similar configuration, however, has no explicit representation of knots as the emergent curvature is an emergent property of the particle interactions. The pattern of the material is thus not constrained by the original configuration pattern, except for the start and end points which are clamped by attractant projection. The spiral curve not only becomes more rounded (Fig. 9, d), but the adaptation and shrinkage proceeds to completely 'unwind' the spiral shape, until a single line connects the start and end points (Fig. 9, i).

The second property relates to the behaviour of the material curve in response to a weak background stimulus, patterned in the shape of the original polyline. Curvature is still dependent on relaxation time but the weak attraction constrains the evolution of the material curve, preferentially detaching from regions with sharp angular changes, whilst remaining attached to longer lines and lines connected by only small changes in angle (This phenomenon can be seen in Fig. 10, and its related online supplementary recording *"Fig. 10: Preferential adhesion to straight paths"*). The material appears to slowly 'peel away' from the corner regions whilst adhering to straight edges.

## 5.2 Computational Properties and Limitations

Comparing the performance of the material computation by the model to classical methods of computing spline curves is somewhat difficult. Classical computation time of spline curves is mainly determined by the number of data points, and thus the number of basis functions to form the curve between the points (see [12] for more details). Computation time in the model (due to the spatial implementation of diffusion within the lattice) is mainly influenced by the area of the lattice containing the original polyline. The number of points does not impact upon the computation time since the entire band of 'material' is considered, instead of discrete points along the polyline. The curve degree does, however, impact the computation time, since degree in the model is related to relaxation time. Fig. 11 shows the relationship between material relaxation and time for a simple triangular shape. The chart plots the distance of the material from the baseline where the material is clamped at each end (Fig. 11, inset images). The circled points in the chart correspond to times when the configuration of the material matches increasing degrees of the approximated spline curve. Note that the time course becomes increasingly non-linear as relaxation progresses.

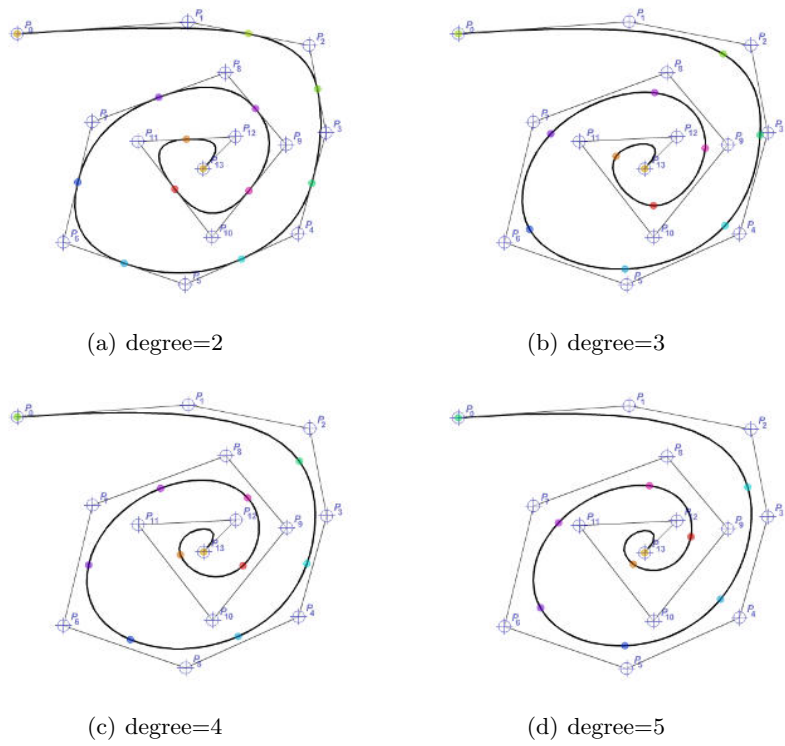

**Fig. 8.** Clamped B-spline of differing degrees, composed of 14 points, clamped at start and end points. Points shown as labelled hollow circles connected by faint lines, knots shown as solid circles on thicker spline curve.

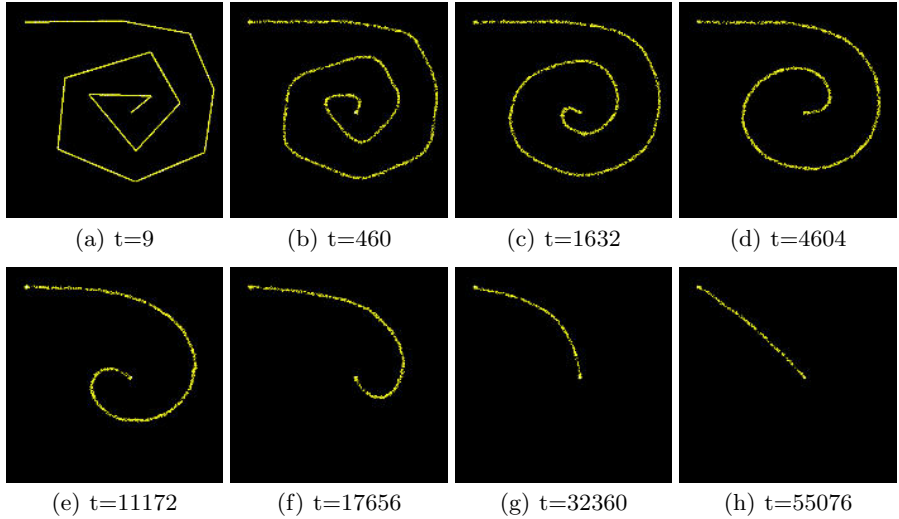

**Fig. 9.** Formation and mechanical evolution of virtual material on spiral configuration. a) material patterned in the original configuration before adaptation, (b-h) evolution of material adaptation clamped at the start and end points by attractant projection.

However, for practical approximation of spline curves, the adaptation occurs well before ($< 5000$ steps) complete material relaxation is required.

A limitation to the approach is that, due to its direct spatial implementation (as opposed to the abstract mathematical representation traditionally used), the material cannot approximate curves with crossing data. In Fig. 12 we show the example of a B-spline curve where the polyline used to generate the curve crosses its own path (Fig. 12,a), which results in spline curves with crossings (curves of different degree are shown in Fig. 12,b-e). When the virtual material is initialised on the same polyline (Fig. 12,f) the resulting material evolution shows an elongation of the crossing point, splitting the curve into separate regions which relax separately before later merging again. As can be seen from Fig. 12,g-j this poorly approximates spline curves of increasing degree. This limitation, however, would not apply to 1D temporal datasets (where it is not possible to 'loop backwards' to previous data) and may not be commonly encountered in 3D datasets (for example for applications of 3D image surface smoothing). This extension of the 2D agent model to a 3D habitat is a logical extension and the subject of ongoing research. Preliminary results suggest that the minimisation and relaxation in seen in 2D data also occurs in 3D surfaces.

## 6 Towards Approximation of Interpolating Spline Curves

Interpolating splines are curves which pass through each data point in the curve. The virtual material method cannot directly approximate interpolating splines when the data points act as attractant projection sites (Fig. 13, a-c, data stimuli indicated at red dots). This is due to the innate minimising behaviour of the virtual material. When the material forms an acute angle where two edges meet (for example the tops of the 'M' shape, Fig. 13, a), the material forms additional points, Steiner nodes, at the vertices where the edges meet (Fig. 13, b). These points move as the material continues to adapt, eventually approximating the Steiner tree (Fig. 13, c), the network connecting all the points which has the minimum distance. Although this is a useful construct for other applications, it is not desirable for the task of approximating spline curves.

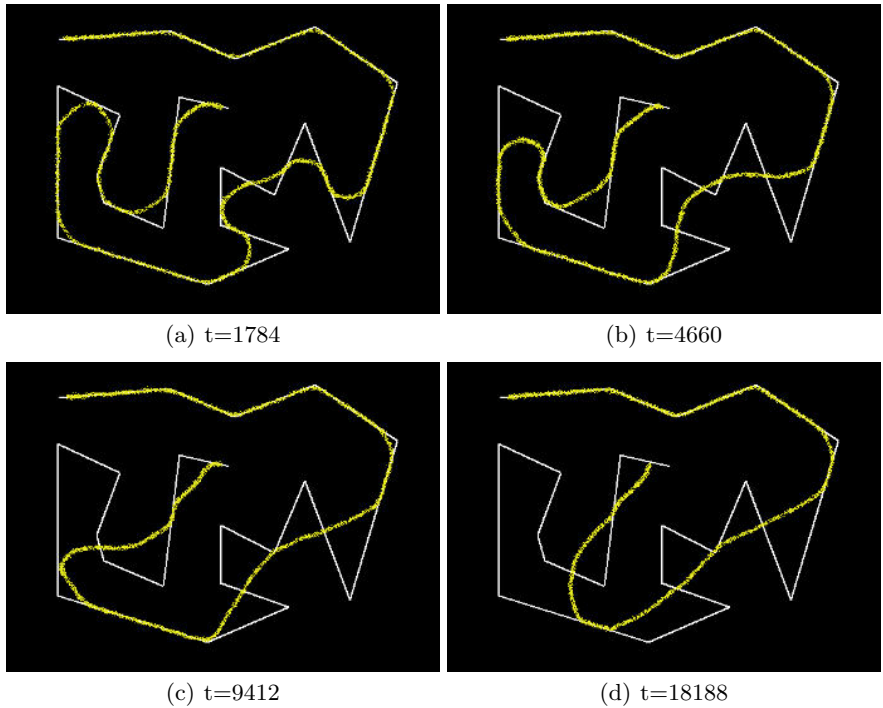

**Fig. 10.** Constraining the material spline via adhesion to weak attractant. A weak attractant signal is projected in the pattern of the white polyline, constraining the evolution of the virtual material. Relaxation constrained by longer lines with shallow angles and more pronounced at short lines and large changes in angle. The material appears to 'peel away' from the original polyline, predominantly at angular regions.

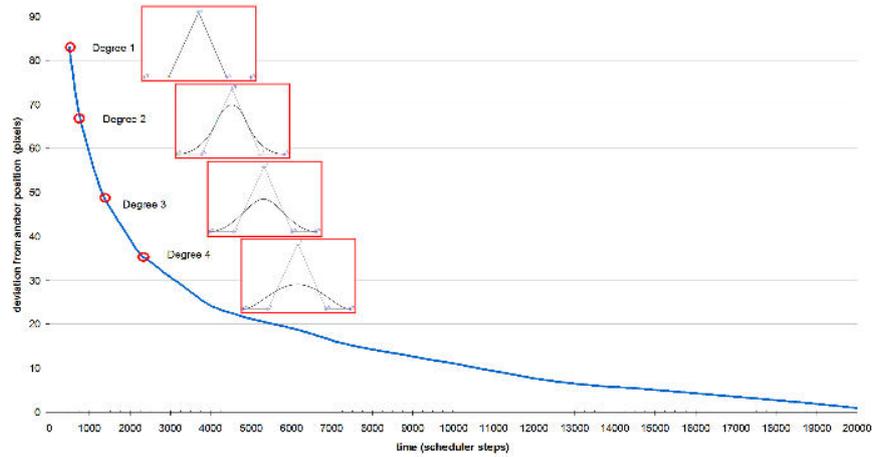

**Fig. 11.** Relationship between virtual material relaxation time and spline curve degree. Chart shows deviation distance of virtual material from the top of a triangle to its base as it relaxes over time (material is clamped at the two end points). Equivalent degree of B-spline curve is indicated by circle positions and inset figures.

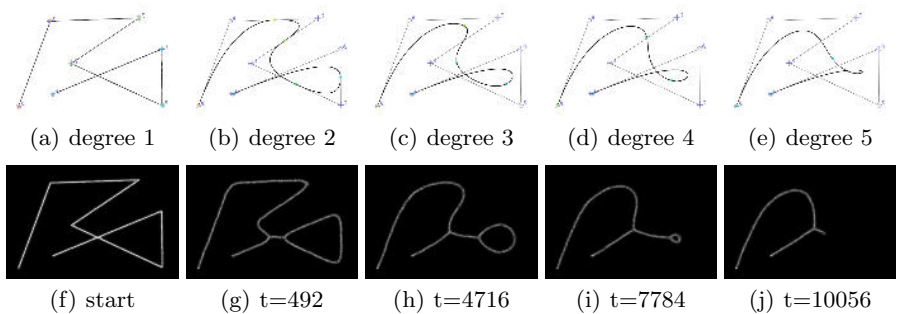

(a) degree 1  (b) degree 2  (c) degree 3  (d) degree 4  (e) degree 5

(f) start  (g) t=492  (h) t=4716  (i) t=7784  (j) t=10056

**Fig. 12.** Material approximation of spline curves is confused by crossing paths. a) original data points and polyline showing crossed paths, b-e) B-spline curves of increasing degree, f) material initialised on original polyline with crossed path, f) separation of loop and curve where paths crossed, h-j) closure of loop and shrinkage of the material poorly approximates spline curves of increasing degree.

### 6.1 Interpolating Splines via Rectilinear Pre-processing

Steiner points only form when the vertex angle is below a certain threshold. This suggests that if the initial stimuli graph could be transformed to give non-acute angles it may be possible for the material to relax and form a curve that passes through all the data points. In Fig. 13, e the initial edges are transformed into rectilinear paths that pass through the nodes. This is achieved via a simple step function where each step occurs midway between the current node and the next node. This ensures that Steiner points will not (initially, at least) be formed at vertices where edges meet. Note that the position of the nodes is not changed at all. When the material is initialised on this rectilinear path and the path stimuli subsequently removed (leaving only the original nodes as stimuli), the material relaxes to approximate an interpolating spline curve. It is important to note, however, that eventually the material will relax to a configuration where the angle at the vertex is sufficiently acute to allow Steiner points to form. Thus, the interpolating spline curve is a transient state in the configuration of the material. Another limitation with the pre-processing approach is the calculation of the rectilinear paths through the nodes. This is trivial to achieve with a 1D dataset (which simplifies the position and direction of the step) but more difficult with 2D datasets.

### 6.2 Interpolating Splines via Interior Containment

The challenge for interpolating splines using a material approach is to allow the material to relax and adapt its shape whilst still forcing the material to pass through the required data points. This can be achieved by initialising the material *inside* the polyline connecting the data points and relaxing within the confines of this shape. The shape is then analogous to a wide pipe containing the material which is anchored at each end. This pipe may be generated from the original polyline by a very simple dilation morphological processing operator to make the polyline thicker (Fig. 14, a and b) and the original data points are represented by the apices of the pipe. The material is initialised within this pipe and anchored at each end point by attractant projection (Fig. 14, c). The material relaxes and shrinks, forming a shorter and smoother path between the two end points and interpolating between the remaining data points (Fig. 14, d).

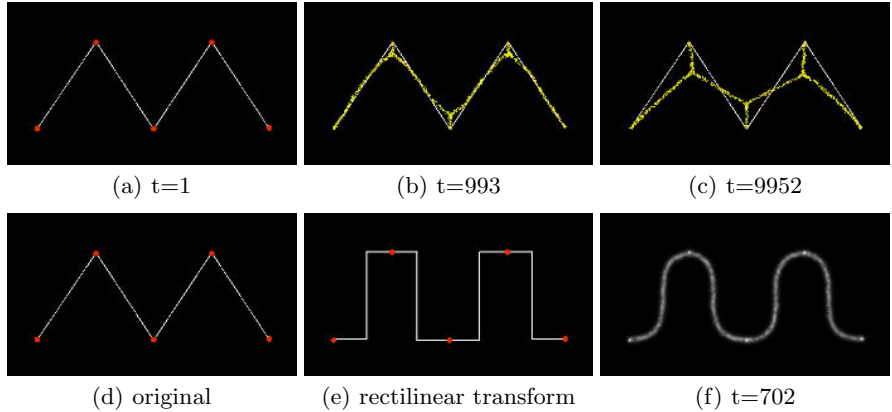

**Fig. 13.** Approximation of interpolating splines by pre-processing the initial configuration stimuli. a) initial M shape stimuli showing source node projection sites (red dots) connected by lines, b) interpolating splines cannot be directly approximated because Steiner points form at the vertices where edges meet, c) the material continues to relax, approximating the Steiner tree, d) original configuration, e) pre-processing of (d) using a simple step function to give rectilinear path through original nodes, f) relaxation of material initialised on rectilinear path approximates an interpolating spline curve.

## 7 Conclusions

True slime mould *Physarum polycephalum* is known to approximate complex spatially represented computation despite it lacking any organised nervous system. The distributed nature of its computation suggests practical applications for self-organised computing collectives. In this article we consider some limitations to spatially implemented unconventional computing by morphological adaptation in slime mould *Physarum polycephalum*, specifically those caused by its adhesion to its substrate. We used a previous particle model of *Physarum* which exhibits deformable network adaptation to try answer the question of 'what would happen if such adhesion was not an obstacle to adaptation?'. The results suggest that distributed computation by simple entities may be useful for data smoothing and image processing applications.

By patterning 1D datasets as variations in the Y-axis along a time path represented by X-axis positions we initialised the particle model along example datasets. The datasets were presented to the model as projected nutrient 'chemoattractant' and we then examined the material evolution under different stimuli conditions. We found that the material approximated the moving average when the stimulus was completely removed and the low-pass filter when stimuli was partially removed.

By directly patterning the shape of 2D datasets (data points connected by polylines) into the model lattice we found that the adaptation of the material approximated spline curves under clamped and unclamped conditions. The degree parameter of spline curves corresponded to the time evolution of the material. B-splines (a type of approximating spline curve) were reproduced by removal or weakening of the original stimulus polyline. Interpolating splines (curves which pass through the data points) were found to be a more difficult problem due to the tendency of the model to form Steiner points at the vertices where acute edges met. Pre-processing of the original dataset was necessary and approximation by internal confinement was the simpler of the two approaches considered. The emergent mechanical properties of the model resulted in some unusual additional features (a mechanical 'unwinding' of the curve to its control points, and a preferential adhesion to longer, straighter paths under weak stimuli).

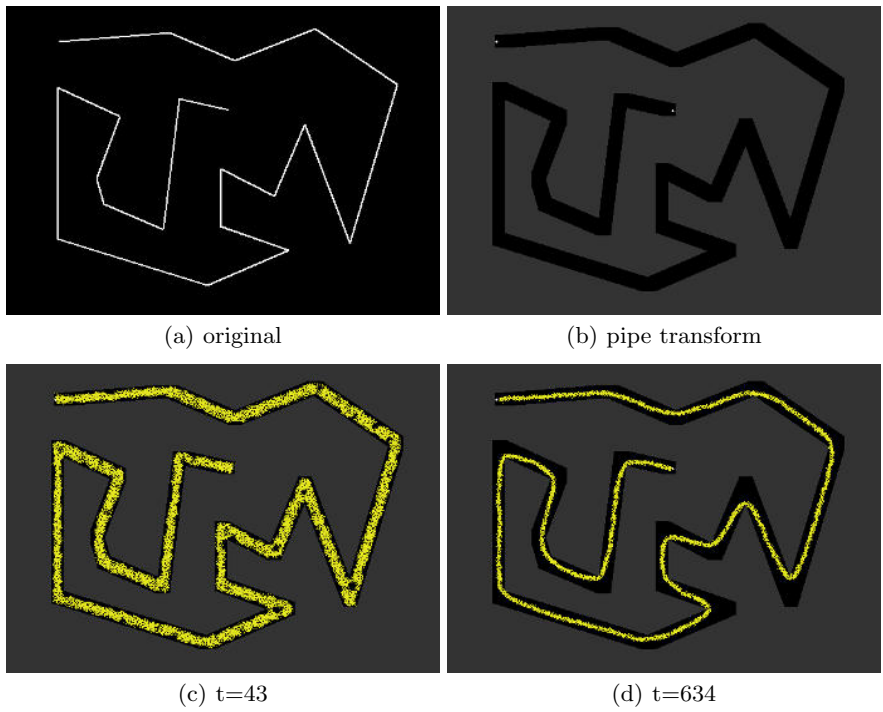

**Fig. 14.** Approximating interpolating splines by interior containment. a) Original polyline connecting data points, b) expansion of original line into wide pipe by simple dilation morphological operator, c) initialisation of material into pipe, d) relaxation, adaptation and shrinkage to form interpolating spline within the confining pipe.

It should be acknowledged that there are already many methods in signal processing which are either inspired by (or directly model by numerical methods) physical processes, including anisotropic diffusion [32], deformable templates [23], and — of course — spline curves themselves. The specific contribution of the approach outlined in this article is that the quasi-physical properties of the virtual material (relaxation, adaptation and shape minimisation) emerge from the very simple particle interactions and the adaptation properties are — as with real slime mould — distributed within the 'material' itself. The material behaviour can be influenced by very simple point attractant stimuli (e.g. clamping points) and edge attractant stimuli (adhesion edges). The method may thus be seen as a minimal complexity example of physically inspired data processing.

There is also possible scope for influence by repellent stimuli: in [22] repellent stimuli have been used to deform Voronoi cell boundaries to generate hybrid Voronoi diagram constructs in which the innate contractile nature of the minimising networks were constrained by the diffusion gradient emanating from the repellent stimuli. This suggests possible future methods of dynamic control of smoothing functions. However, the flip side of distributed control is that previously simply specifiable parameters may be transformed into a more nebulous method of control. A good example is the degree of B-spline curvature which is dependent on the amount of relaxation time of the model. Although this is intuitively simple to understand (material left to relax for a longer time will relax to a greater degree) it may be less easy to directly relate (in terms of simple parameters) curvature to relaxation time.

To conclude we have shown that, by considering particular physical limitations of adaptation in slime mould *Physarum polycephalum*, we can explore speculative 'what if?' questions by utilising a particle model of *Physarum* that shares properties of having simple component parts, local interactions and distributed emergent behaviour. Using this model we spatially approximated complex data smoothing and spline curve functions by means of morphological adaptation. This generates both a baseline minimal specification for computation via physical processes and also expands the computational repertoire of unconventional computing devices. Further work could be carried out towards extending the approach into robust deformable template models (for example with contour completion and size adaptation), 3D surface smoothing functions and examining the potential of the approach for related data analysis, combinatorial optimisation and constraint satisfaction problems.

## 8 Appendix: Particle Model Description

We used a multi-agent approach to generate the *Physarum*-like behaviour. This approach was chosen specifically specifically because we wanted to reproduce the generation of complex behaviour using very simple component parts and interactions, and no special or critical component parts to generate the emergent behaviour. Although other modelling approaches, notably cellular automata, also share these properties, the direct mobile behaviour of the agent particles renders it more suitable to reproduce the flux within the plasmodium. The multi-agent particle model of *Physarum* used to generate the relaxation and adaptation [18] uses a population of coupled mobile particles with very simple behaviours, residing within a 2D diffusive lattice. The lattice stores particle positions and the concentration of a local diffusive factor referred to generically as chemoattractant. Particles deposit this chemoattractant factor when they move and also sense the local concentration of the chemoattractant during the sensory stage of the particle algorithm. Collective particle positions represent the global pattern of the material. The model runs within a multi-agent framework running on a Windows PC system. Performance is thus influenced by the speed of the PC running the framework. The particles act independently and iteration of the particle population is performed randomly to avoid any artifacts from sequential ordering.

### 8.1 Generation of Virtual Plasmodium Cohesion and Shape Adaptation

The behaviour of the particles occurs in two distinct stages, the sensory stage and the motor stage. In the sensory stage, the particles sample their local environment using three forward biased sensors whose angle from the forwards position (the sensor angle parameter, SA), and distance (sensor offset, SO) may be parametrically adjusted (Fig. 15a). The offset sensors generate local indirect coupling of sensory inputs and movement to generate the cohesion of the material. The SO distance is measured in pixels and a minimum distance of 3 pixels is required for strong local coupling to occur. For the experiments in this article we used an SO value of 5. Increasing the SO value results in thicker curves, a faster evolution of the adaptation and a coarser approximation of the original spline curves (see supplementary material for examples at increasing SO scales). During the sensory stage each particle changes its orientation to rotate (via the parameter rotation angle, RA) towards the strongest local source of chemoattractant (Fig. 15b). Variations in both SA and RA parameters have been shown to generate a wide range of reaction-diffusion patterns [17] and for these experiments we used SA 90 and RA 45 which results in stronger and more rapid adaptation of the virtual material. After the sensory stage, each particle executes the motor stage and attempts to move forwards in its current orientation (an angle from 0–360 degrees) by a single pixel forwards. Each lattice site may only store a single particle and particles deposit chemoattractant into the lattice (5 units per step) only in the event of a successful forwards movement. If the next chosen site is already occupied by another particle move is abandoned and the particle selects a new randomly chosen direction.

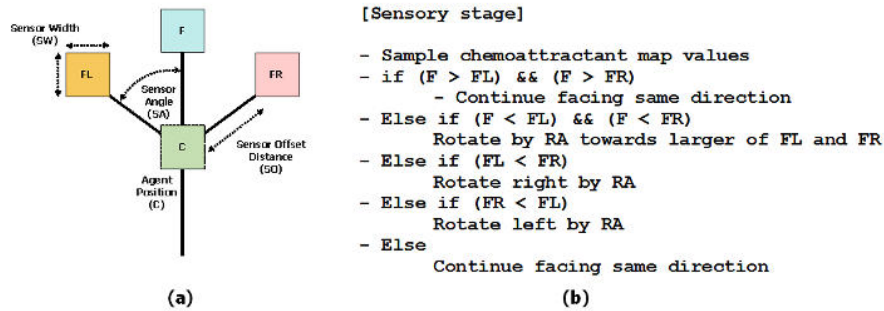

**Fig. 15.** Architecture of a single particle of the virtual material and its sensory algorithm. (a) Morphology showing agent position 'C' and offset sensor positions (FL, F, FR), (b) Algorithm for particle sensory stage.

### 8.2 Problem Data Representation

The spatially implemented computation requires that the data configuration be contained within the 2D lattice containing the particle population. Data configurations are loaded as greyscale image files and this data is interpreted by the scheduler and projected into the diffusive lattice as a virtual chemoattractant to attract the particle population. For 1D data smoothing the Y-axis represents the actual data values and the X-axis represents the time series. Comparison data smoothing curves (moving average and low-pass filter) were numerically generated using manual construction of filter kernels of increasing width using Microsoft Excel.

For 2D curve approximation B-spline curves and interpolating spline curves were generated using the interactive software method by Foretník [11], allowing the user to specify points. The software then generated the respective polylines and curves with specifiable degree and clamping options. For the material approximation the polylines connecting the data points were projected into the lattice at a value of 2.55 units per data pixel. This projection causes the particle population to be attracted to the data configuration, initially preventing adaptation of the population. The particle population was initialised on the 1D data line or 2D polyline respectively and particle movement was halted for short time (20 scheduler steps) to allow attraction to the initial data stimuli. To initiate complete relaxation the source data was removed from the lattice and any clamping points represented by projection of discrete data points at 2.55 units per pixel. To initiate partial relaxation due to weakened attractant stimuli the data projection was weakened to 0.255 units per data pixel, allowing relaxation, but constraining the relaxation along straighter paths.

### 8.3 Material Shrinkage Mechanism

Relaxation and adaptation of the virtual material is implemented via tests executed at regular intervals as follows. If there are 1 to 10 particles in a $9 \times 9$ neighbourhood of a particle, and the particle has moved forwards successfully, the particle attempts to divide into two if there is a space available at a randomly selected empty location in the immediate $3 \times 3$ neighbourhood surrounding the particle. If there are 0 to 24 particles in a $5 \times 5$ neighbourhood of a particle the particle survives, otherwise it is deleted. Deletion of a particle leaves a vacant space at this location which is filled by nearby particles, causing the collective to shrink slightly. As the process continues the material shrinks and adapts its morphology to the stimuli provided by the polyline or clamping points. If no external stimuli are present the material will eventually adapt to a minimal circular shape and shrink down to a small cluster of points in size. The frequency at which the growth and shrinkage of the population is executed determines a turnover rate for the particles. The frequency of testing for particle division and particle removal was every 2 scheduler steps. This relatively high frequency (compared to other applications using the virtual material approach, e.g. [21]) is due to the strong shrinkage invoked by the particular SA/RA combination used, necessitating a high adaptation frequency to maintain connectivity of the band of material as it adapts and shrinks.

## Acknowledgements

This work was supported by the EU research project "Physarum Chip: Growing Computers from Slime Mould" (FP7 ICT Ref 316366)